\documentclass[a4paper,10pt,conference]{IEEEtran}
\usepackage{graphicx}
\usepackage{balance}
\ifCLASSINFOpdf
\else
\fi
\begin{document}
%
\title{To Use or Not to Use: CPUs' Cache Optimization Techniques on GPGPUs}

\author{\IEEEauthorblockN{D.R.V.L.B. Thambawita}
\IEEEauthorblockA{Department of Computer Science and Technology\\
Uva Wellassa University\\
Badulla, Sri Lanka\\
Email: vlbthambawita@gmail.com}
\and
\IEEEauthorblockN{Roshan G. Ragel and Dhammike Elkaduwe}
\IEEEauthorblockA{Department of Computer Engineering\\
University of Peradeniya\\
Peradeniya, Sri Lanka\\
Email: [ragelrg, dhammika.elkaduwe]@gmail.com}
}


%



\maketitle

\begin{abstract}
General Purpose Graphic Processing Unit(GPGPU) is used widely for achieving high performance or high throughput in parallel programming. This capability of GPGPUs is very famous in the new era and mostly used for scientific computing which requires more processing power than normal personal computers. Therefore, most of the programmers, researchers and industry use this new concept for their work. However, achieving high-performance or high-throughput using GPGPUs are not an easy task compared with conventional programming concepts in the CPU side. In this research, the CPU’s cache memory optimization techniques have been adopted to the GPGPU’s cache memory to identify rare performance improvement techniques compared to GPGPU's best practices.  The cache optimization techniques of blocking, loop fusion, array merging and array transpose were tested on GPGPUs for finding suitability of these techniques. Finally, we identified that some of the CPU cache optimization techniques go well with the cache memory system of the GPGPU and shows performance improvements while some others show the opposite effect on the GPGPUs compared with the CPUs.

\textit{Keywords - GPGPU, CPU, Cache Optimization, CUDA, Fermi Architecture}
\end{abstract}


%
\IEEEpeerreviewmaketitle

\section{Introduction}
Today, parallel computing with GPGPUs is one of the most popular research areas in high-performance computing. Doing parallel computing more accurately, quickly and cost effectively is the main benefit for  researchers by using GPGPUs. Therefore, our previous research article discusses that when the GPGPUs are suitable for running a program compared with CPUs \cite{DBLP:journals/corr/ThambawitaRE14}.
Many numbers of research have been done for CPUs' cache optimizations. On the other hand, NVIDIA introduced several techniques for improving the performance of GPGPU computing by providing various memory improvement techniques \cite{manual_cuda_c_best_practice}. However, it doesn't contain GPGPU cache optimizations and effects of cache usage of GPGPUs. Therefore, there is a lack of information related to the cache optimization techniques of GPGPUs.

This research is about that how could the GPGPU's memory usage be optimized using cache optimization techniques. There are several common cache optimization techniques which are based on the CPUs' cache architectures like applying the blocking technique, applying the loop fusion, merging arrays and applying array transpose. These techniques have been modified and adapted to the GPGPUs for testing the performance improvements. While these CPU cache optimization techniques were adopted into the GPGPU, the cache parameters of the GPGPU have been changed for measuring the effects of changing the cache parameters for the performance.

\section{Related work}
Hennessy and Patterson \cite{book_computer_architecture} have listed six basic cache optimizations and they are: (1) larger block size to reduce miss rate (2) bigger caches to reduce miss rate; (3) higher associativity to reduce miss rate; (4) multilevel caches to reduce miss penalty; (5) giving priority to read misses overwrites to reduce miss penalty; and (6) avoiding address translation during indexing of the cache to reduce hit time. However, the main CPU cache optimization techniques based on programming side have been categorized like data access optimizations and data layout optimizations in the lecture notes written by Markus Kowarschik and Christian Weiß \cite{Kowarschik03anoverview} and our experiments are based on these techniques.

Chen et al.\cite{Xuhao_Chen_et_al} have proposed a new specialized cache management policy for GPGPUs. CPU cache architecture is designed especially for reducing memory latency while GPGPU cache architecture is designed for achieving high throughput. However, the existing CPU cache optimization techniques were not used in their research because of limitations of the CPU cache optimization techniques over the GPGPU. These limitations were not explained to remove the CPU cache optimization techniques for GPGPUs in this research. Less effectiveness of GPGPUs' cache was discussed  in this paper and it helped us for understanding the behavior of GPGPU cache.

Mahmoud Khairy, Mohamed Zahran, and Amr G. Wassal \cite{Mahmoud_Khairy_et_al} have experimented three techniques for utilizing GPGPU cache efficiently. They are,
\begin{itemize}
	\item Dynamically Bypassing Streaming Applications
	\item Dynamic Warp Throttling via Cores Sampling (DWT-CS)
	\item Pseudo Random Interleaved Cache (PRIC)
\end{itemize}
\IEEEpubidadjcol
These techniques are closer to the prior discussed paper's technique \cite{Xuhao_Chen_et_al}. They used GPGPU-Sim \cite{GPU_simulator} for all the experiments. A limitation of this research is, experimental results and conclusions depend on a virtual system and it cannot be trusted on real GPGPUs. However, this research is not based on adopting CPU cache optimization techniques into GPGPU cache optimization techniques.

Meltzer R., Zeng C., and Cecka C. \cite{gpuCache_meltzer2013micro} have done some experiments for identifying cache structure of a GPGPU architecture. They have used c2070 Tesla card for experiments. It is also based on Fermi architecture and it is similar to our c2075 GPGPU. Confirming cache sizes and determining associativity of L1 and L2 cache have been done as the main part of this research. Therefore, this research helps us to identify cache architecture of the GPGPU. More information about L1, L2, TLB structure, texture memory structure, global memory throughput,  global memory latency, shared memory throughput and latency were collected from a research done by Mei et al. \cite{gpuCache_Xinxin_Mei}.

However, we adopted selected CPU cache optimization techniques to the GPGPU architecture for identifying the effects of them. Then, possibilities of adopting CPU cache optimizations over the GPGPU for gaining the better performance were discussed here.

\section{Background knowledge}
\subsection{CPU and GPGPU cache architectures}
\begin{figure}[!t]
	\centering
	\begin{minipage}[b]{0.23\textwidth}
		\includegraphics[width=\textwidth]{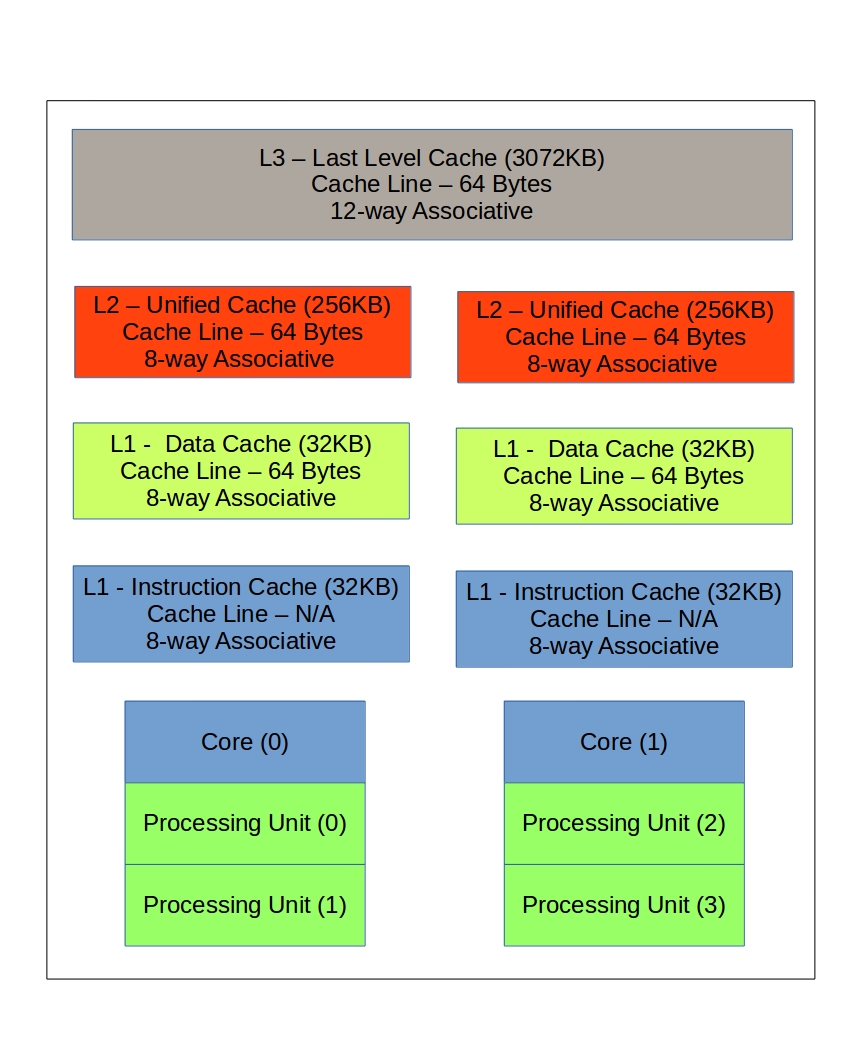}
	\end{minipage}
	\hfill
	\begin{minipage}[b]{0.23\textwidth}
		\includegraphics[width=\textwidth]{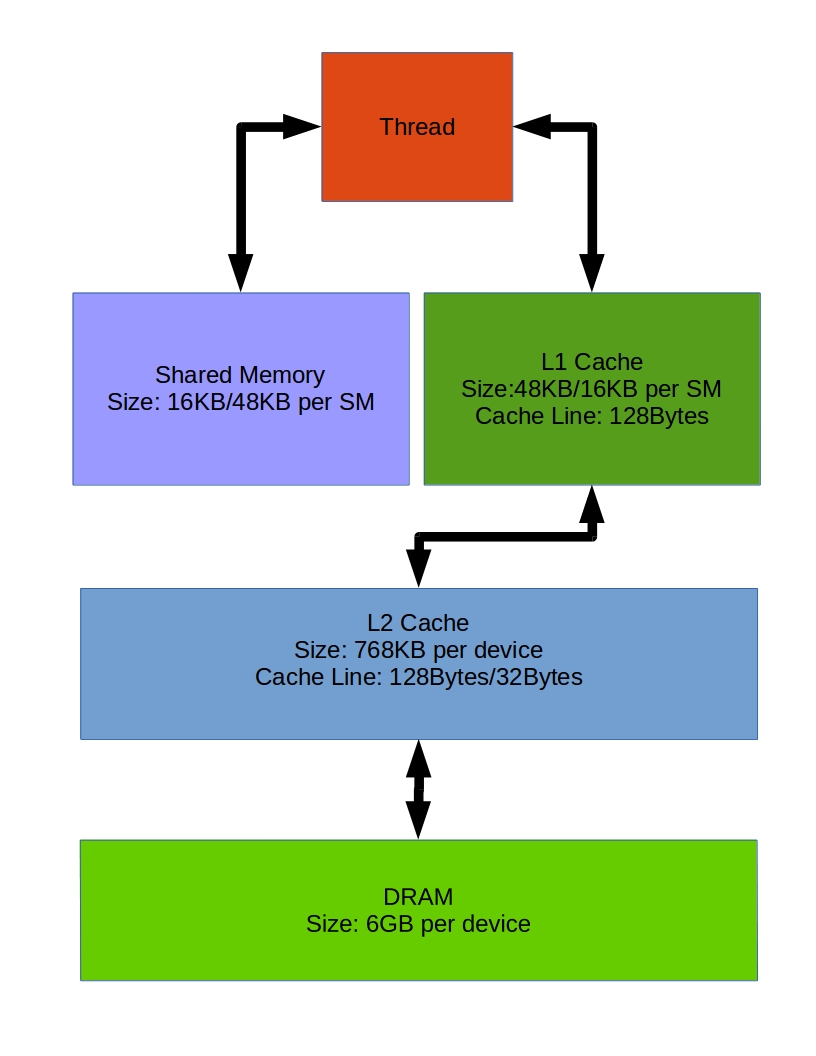}
	\end{minipage}
	\caption{Cache memory hierarchy of CPUs and GPGPUs (Fermi architecture).}
	\label{fig:cpu_and_GPGPU_cache_architecture}
\end{figure}

Our main purpose of this research is adopting CPU cache optimization techniques to the GPGPU. Therefore, it is required to know differences between CPU cache and GPGPU cache architecture. Fig. \ref{fig:cpu_and_GPGPU_cache_architecture} shows the cache architecture of CPUs and GPGPUs.

Nowadays, most CPUs have three-level cache architecture while GPGPUs are consisted with two-level cache architecture. However, CPUs' cache architecture is well established compared with experimental GPGPUs' cache architecture. GPGPUs' cache can be configured by the user as required for the application. Main configurations are, L1 cache can be disabled, L1 cache size can be changed to different sizes like 48KB and 16KB corresponding to 16KB and 48KB shared memory. If we disabled the L1 cache, this architecture of the GPGPU will change the cache line size from 128bytes to 32bytes.

\subsection{CPU cache optimization techniques}

Within this paper, blocking technique and loop fusion technique are discussed under the data access optimization techniques while  array merging and array transpose are discussed under data layout optimizations.

Blocking or loop tiling is a technique to divide large data set into small blocks which are fitted to the cache lines of the cache architecture for improving the performance. This will improve temporal locality and spatial locality for reducing cache misses. The idea behind this is minimizing cache line replacements.

Loop fusion is combining two loops together. For this, both loops should have the same number of iterations and some common variables. Then it reduces the number of iterations to the half and improves the temporal locality. On the other hand, it reduces the number of branching conditions in fused loops.

Array merging is a technique for joining two or more arrays together to get better performance. The spatial locality can be improved as the main advantage. In addition to this, cross interferences can be reduced in some cases because these merged arrays are an alternative structure instead of original data structures which have cross interferences.

The transposing array technique is interchanging dimensions of a two-dimensional arrays. It means, if array ``a" is a[x][y], then the transposed array is a[y][x]. The effect of this technique is similar to the effect of the interchanging loops. While the interchanging loops changes the instruction order, the transposing arrays changes the order of the data structure. Like in interchanging loops, the main idea behind this is changing the stride access number. As a result, it improves the spatial locality.

\section{Methodology}

\subsection{Method for testing the performance of blocking technique}
The blocking technique was analyzed using the matrix multiplication. Two matrices A and B are multiplied and answer C is calculated. This sample program has been run on CPU without applying blocking technique and with two different blocking techniques from two different sources \cite{book_computer_architecture} \cite{manual_cuda_c_programming_guid}.

\begin{figure}[!t]
	\centering
	\begin{minipage}[b]{0.23\textwidth}
		\includegraphics[width=\textwidth]{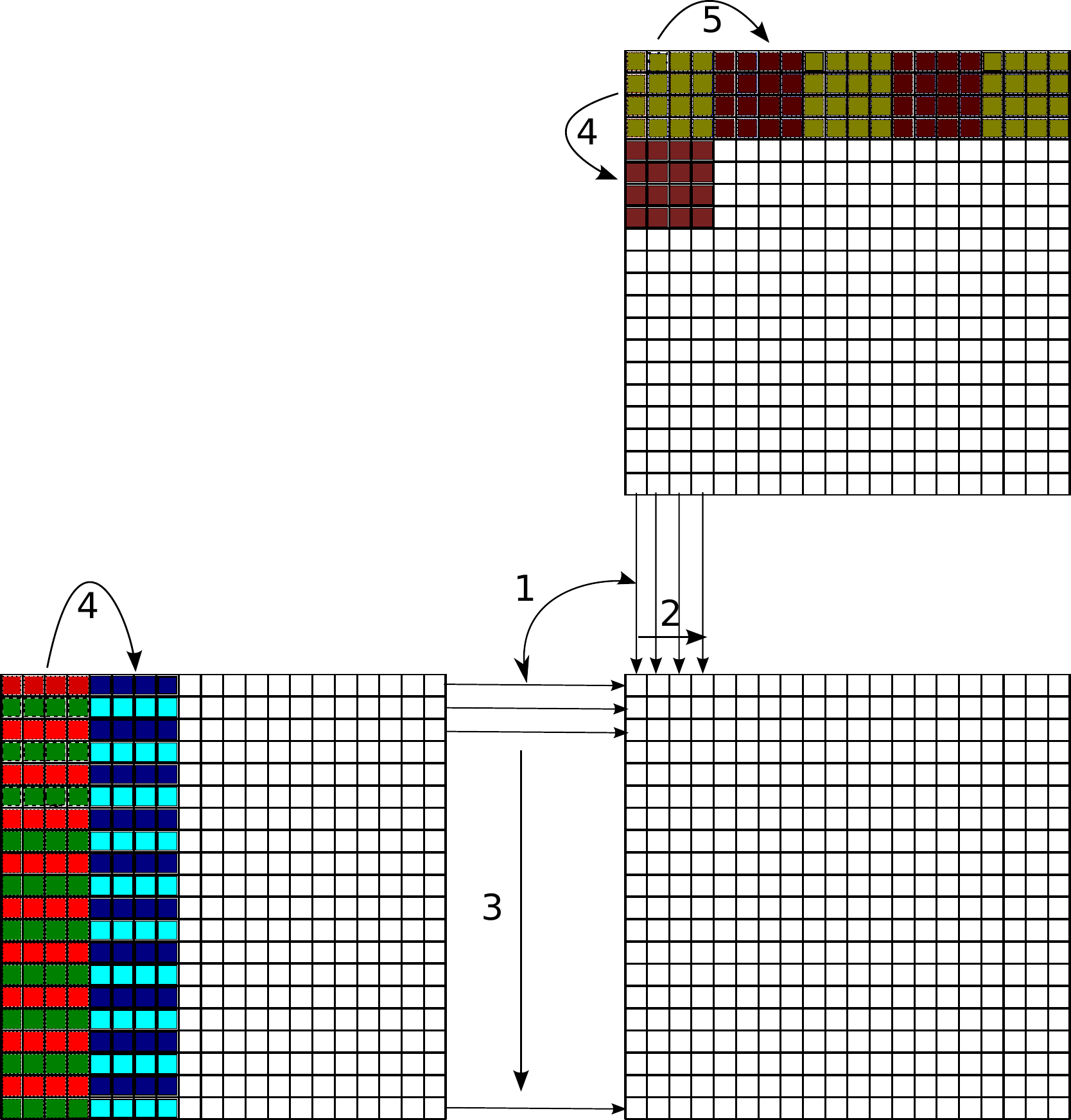}
	\end{minipage}
	\hfill
	\begin{minipage}[b]{0.23\textwidth}
		\includegraphics[width=\textwidth]{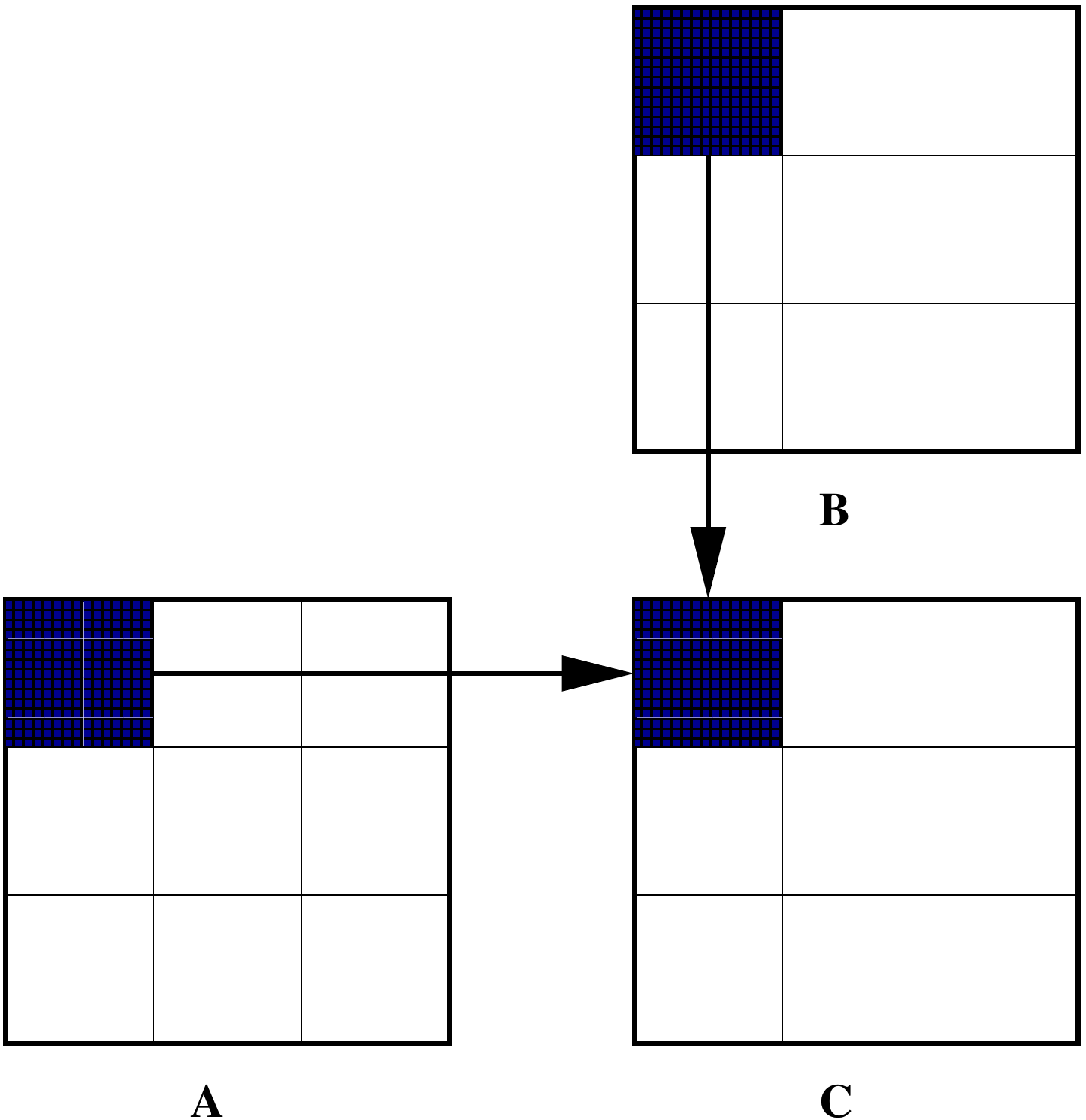}
	\end{minipage}
	\caption{Two different blocking techniques from two different sources \cite{book_computer_architecture} \cite{manual_cuda_c_programming_guid}. First technique uses small blocks from the first matrix and large blocks from the second matrix. Second method uses equal size blocks form both matrices.}
	\label{fig:blocking_techniques}
\end{figure}

Behaviors of these two techniques are illustrated in Fig. \ref{fig:blocking_techniques}. However, one method was selected for GPGPUs based on the performance. When the blocking technique was tested with the GPGPU, it was adopted to different cache memories and cache configurations of the GPGPU for comparing the performance differences. The first experiment has been done with disabled L1. The second experiment was based on default configuration of the cache memory of the GPGPU. Final test was done with the largest L1 cache of the GPGPU.

\subsection{Method for testing the performance of loop fusion technique}
The main CPU test for loop fusion has been done by using two separate loops with common memory accesses. It is required to match the number of branching conditions in both fused and non-fused loops. If the number of branches in two different methods is not equal, then it is very difficult to identify the main factor for changing the performance. Then loop unrolling technique was used for matching a number of branching conditions. Then, common variables within for loops have been used in this experiment for making some common factors between two loops. The GPGPU adaptation has been done after getting the main idea about effects of loop fusion on the CPU side.

However, the loops within the GPGPU are kernels because a loop is running on a GPGPU in parallel. Therefore, kernel fusion is the technique in GPGPUs corresponding to the loop fusion in CPUs. Here, the first experiment of the GPGPU side has been done with default cache configurations of the GPGPU. However, two common memory accesses are included within kernels also. After identify the main effect of the kernel fusion on the GPGPU, cache configurations of the GPGPU have been changed for measuring effects of the cache configurations to the kernel fusion.

\subsection{Method for testing the performance of array merging technique}
\begin{figure}[!t]
	\centering
	\includegraphics[width=3.3in]{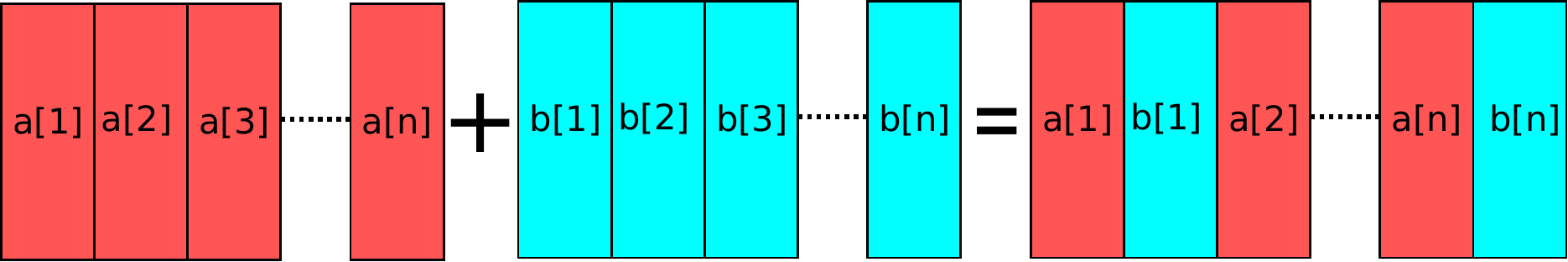}
	\caption{Basic idea behind the array merging technique.}
	\label{fig:array_merging}
\end{figure}

As the array merging test case, three arrays were initiated with the size of nxn. Then, the main for loop is accessing these three arrays to modify the content. An element of the array is accessed by skipping 16 elements from the previous access because this size is equal to the 16 x 8 bytes = 128 bytes. This amount is twice as the size of the L1 cache line of the CPU. The main target of applying this mechanism is accessing elements of arrays using new cache lines. Without merging technique, data elements from three different arrays were accessed from three different locations from the main memory of the CPU. After applying array merging, less main memory accesses can be expected due to spatial locality of the CPU cache. This array merging idea can be identified easily using the Fig. \ref{fig:array_merging}.

In the GPGPU test, the kernel accesses the arrays’ elements from the global memory through the L2 and L1 cache of the GPGPU like the CPU is accessing from the main memory through the L3, L2 and L1 cache. As the main adaptation, the index value is changed 64 by 64 for each thread to make different cache line accesses for consequent memory accesses on the global memory. Then, cache configurations were changed one by one for measuring the effect of cache configurations for the array merging.

Next, texture memory of the GPGPU is identified as another suitable memory location for applying this array merging technique. This texture memory is designed specially to handle 2D data structures within the GPGPU. Next experiment was done to measure the effect of array merging technique within this texture memory of the GPGPU by applying the array merging into two different texture memory arrays.

\subsection{Method for testing the performance of array transpose}
The matrix multiplication is an example for testing the effect of array transpose. Therefore, first experiment has been done within the CPU by applying array transpose technique and without applying array transpose. Then, result was analyzed to get an idea about effects of array transpose in CPU side.

\begin{figure}[!t]
	\centering
	\begin{minipage}[b]{0.23\textwidth}
		\includegraphics[width=\textwidth]{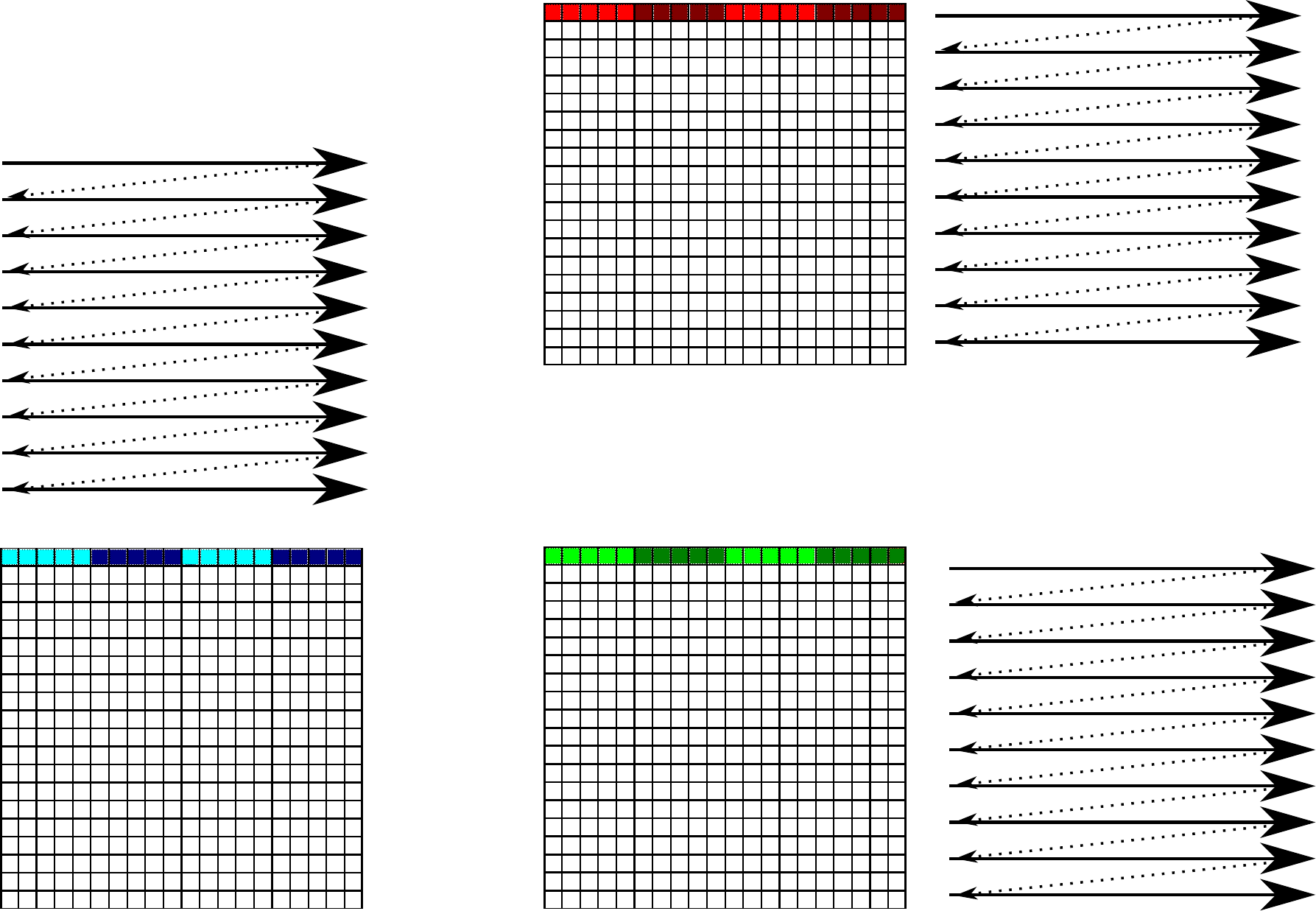}
	\end{minipage}
	\hfill
	\begin{minipage}[b]{0.23\textwidth}
		\includegraphics[width=\textwidth]{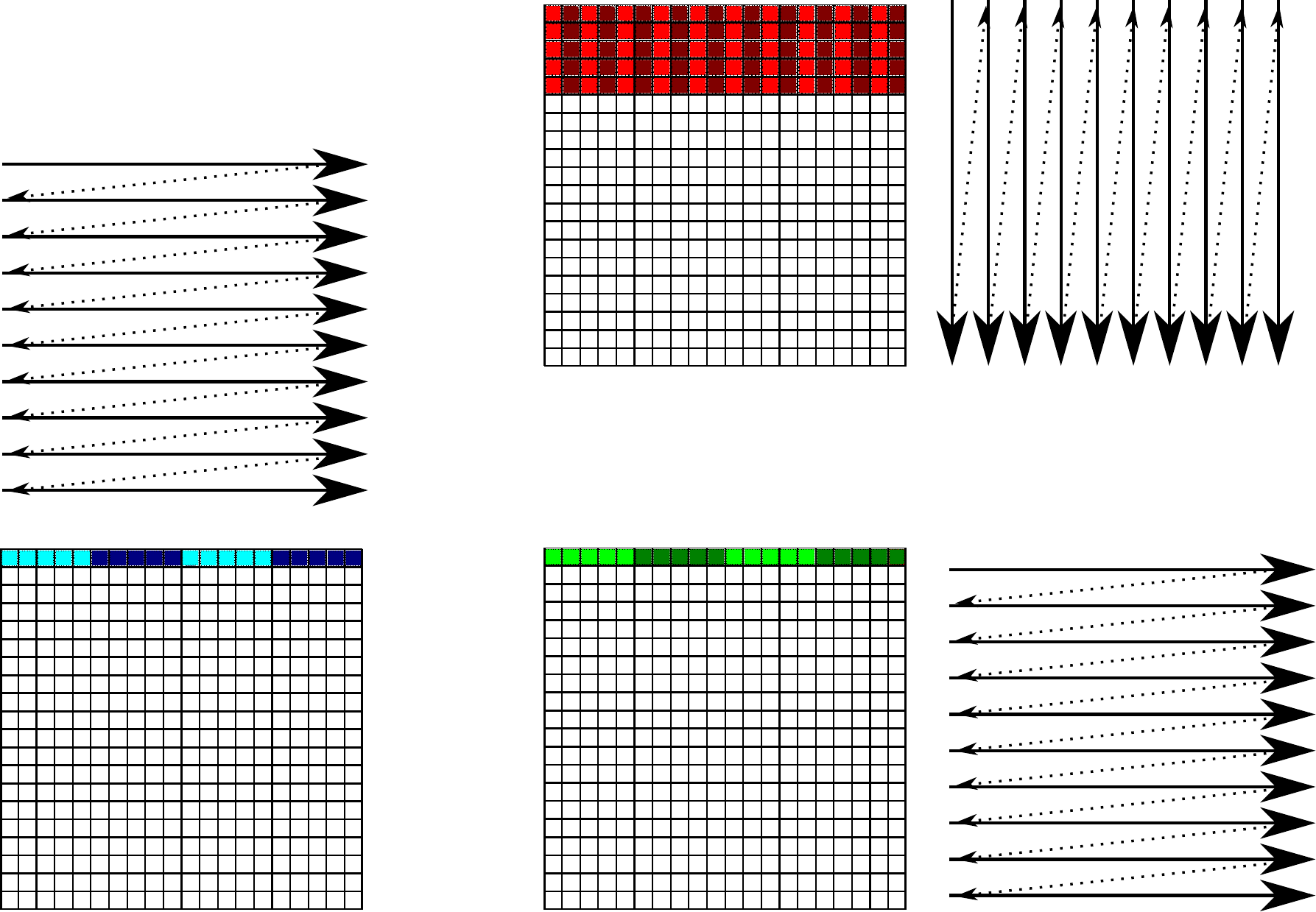}
	\end{minipage}
	\caption{Basic matrix multiplication is shown in first figure while second figure is illustrated that how to use transposed matrix for matrix multiplication.}
	\label{fig:matrix_mul_with_transpose}
\end{figure}

Within the GPGPU test, each element of the answer matrix was calculated by an individual thread of the grid. Therefore, consecutive elements of the matrix will not be accessed by the same thread. However, transposed matrix has been manipulated from the host before passing it into the device. This transposing method is a host code and it is not running in parallel. Normal matrix multiplication process and matrix multiplication with transposed matrix can be seen at Fig. \ref{fig:matrix_mul_with_transpose}.

\section{Cache architectures of the experimental setups}
The cache structure of the CPU side of the experimental setup has been identified using an Intel architecture optimization manual [3] and an article about MESI protocol written by Gomez-Luna et al.[4]. This information tabulated in Table \ref{tbl:cpu_cache_architecture}.
The cache memory information about the GPGPU of the experimental setup was gathered from a CUDA programming book written by S. Cook [5] and the manual of Fermi architecture [6]. The details of GPGPU cache architecture have been given in Table \ref{tbl:cache_architecture_GPGPU}.

\begin{table}[!t]
	\centering
	\caption{Intel Core (TM) i5 – 3230M CPU 2.6 GHz – Ivy Bridge micro-architecture with 8GB RAM}
	\label{tbl:cpu_cache_architecture}
	\begin{tabular}{lllll}
		\hline
		& \begin{tabular}[c]{@{}l@{}}Cache \\ size\end{tabular} & \begin{tabular}[c]{@{}l@{}}Cache \\ line size\end{tabular} & Associativity & Description \\
		\hline
		L1 cache & 32KB & 64bytes & 8-way &  \\
		L2 cache & 256KB & 64bytes & 8-way &  \\
		L3 cache & 3072KB & 64bytes & 12-way & Shared Memory \\
		\hline
	\end{tabular}
\end{table}

\begin{table}[!t]
	\centering
	\caption{Tesla c2075 GPGPU's cache architecture-6GB global memory}
	\label{tbl:cache_architecture_GPGPU}
	\begin{tabular}{lllll}
		\hline
		& \begin{tabular}[c]{@{}l@{}}Cache \\ size\end{tabular} & \begin{tabular}[c]{@{}l@{}}Cache \\ line size\end{tabular} & Associativity & Description \\
		\hline
		L1 cache & \begin{tabular}[c]{@{}l@{}}48KB/\\ 16KB\end{tabular} & 128bytes & Not mentioned & \begin{tabular}[c]{@{}l@{}}can be disable\\ by using\\ -Xptxas-dlcm=cg\\ compile flag\end{tabular} \\
		\begin{tabular}[c]{@{}l@{}}Shared\\ memory\end{tabular} & \begin{tabular}[c]{@{}l@{}}16KB/\\ 48KB\end{tabular} & 128bytes & Not mentioned & \begin{tabular}[c]{@{}l@{}}can be used \\ manually\end{tabular} \\
		L2 cache & 768KB & \begin{tabular}[c]{@{}l@{}}128bytes/\\ 32bytes\end{tabular} & Not mentioned & Unified cache \\
		\hline
	\end{tabular}
\end{table}

\section{Results and discussion}
Main results obtained from experiments are discussed here with corresponding graphs. This section is divided into subsections for each cache optimization technique discussed in the methodology section.

\subsection{Effect of blocking techniques}
\begin{figure}[!t]
	\centering
	\includegraphics[width=3.3in]{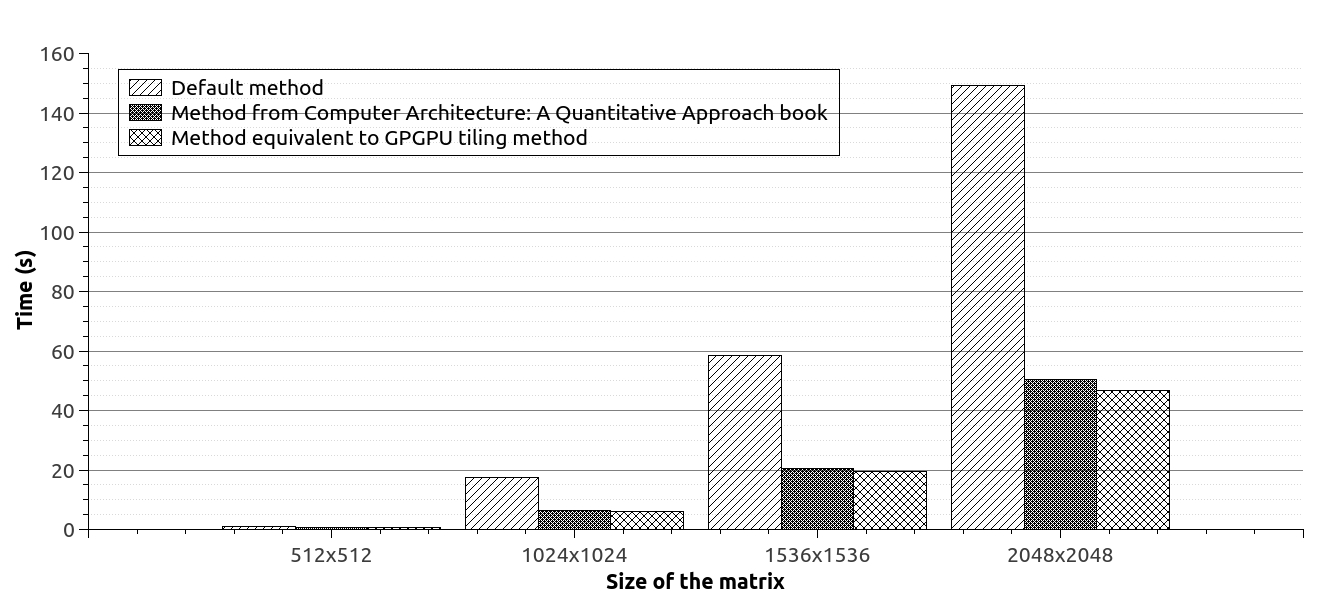}
	\caption{Effect of blocking (tilling) techniques on the CPU.}
	\label{fig:cpu_blocking}
\end{figure}

Within the above Fig. \ref{fig:cpu_blocking}, it is easy to identify a clear difference between the performance of non-blocking and blocking matrix multiplication over the CPU. Blocking technique shows big performance improvement over the non-blocking technique of the CPU. The reason for this performance gain is mapping data into cache perfectly due to these blocks. The second blocking method discussed in CUDA programming manual \cite{manual_cuda_c_programming_guid} shows better performance in the CPU than the technique discussed in the computer architecture book \cite{book_computer_architecture}.

\begin{figure}[!t]
	\centering
	\includegraphics[width=3.3in]{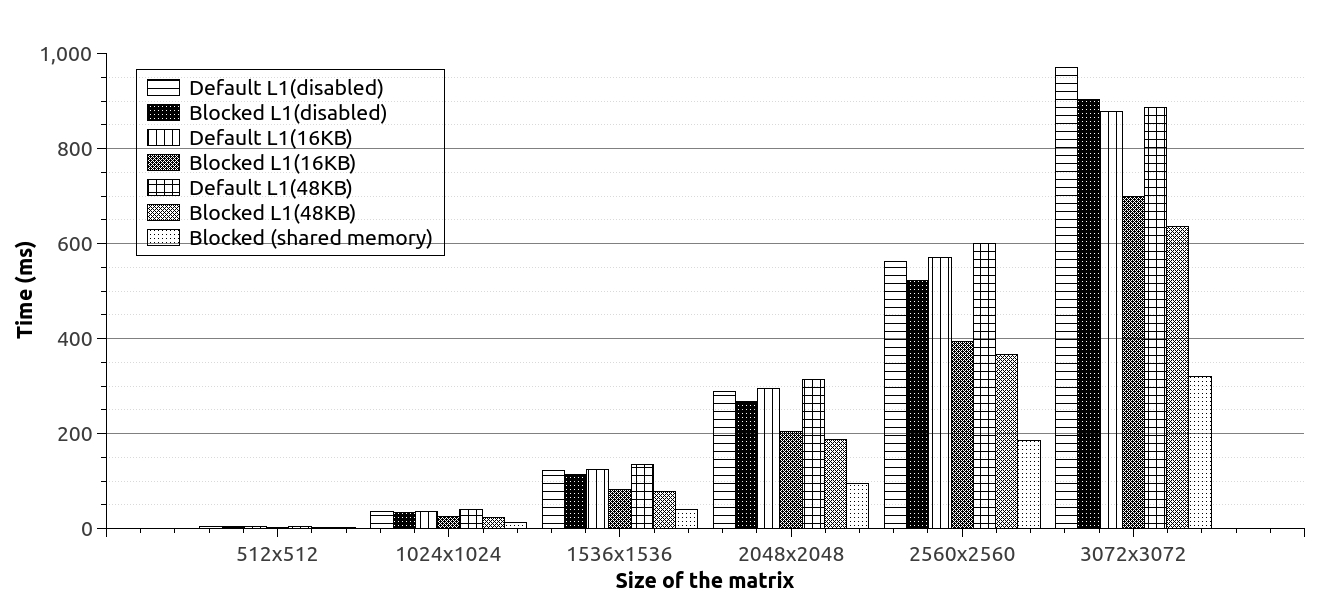}
	\caption{Effect of the blocking technique on Fermi GPGPU with various cache configurations.}
	\label{fig:gpgpu_blocking}
\end{figure}

The Fig. \ref{fig:gpgpu_blocking} shows a clear performance differences between default matrix multiplication and the matrix multiplication with blocking technique on the GPGPU. However, best performance could be achieved while the shared memory is used. Other three methods which are used L1 cache and L2 cache shows better performance than default method. However, these three methods show poor performance than shared memory based method. The reason is, shared memory is loaded manually and L1 cache is loaded using predefined algorithms of the GPGPU. Therefore, L1 cache is loaded with unnecessary memory locations also while the shared memory is loaded with only required memory locations.

In this blocking technique, big cache line size shows better performance than small cache line size because the large portion from a block can be held.

\subsection{Effect of loop fusion}

\begin{figure}[!t]
	\centering
	\includegraphics[width=3.3in]{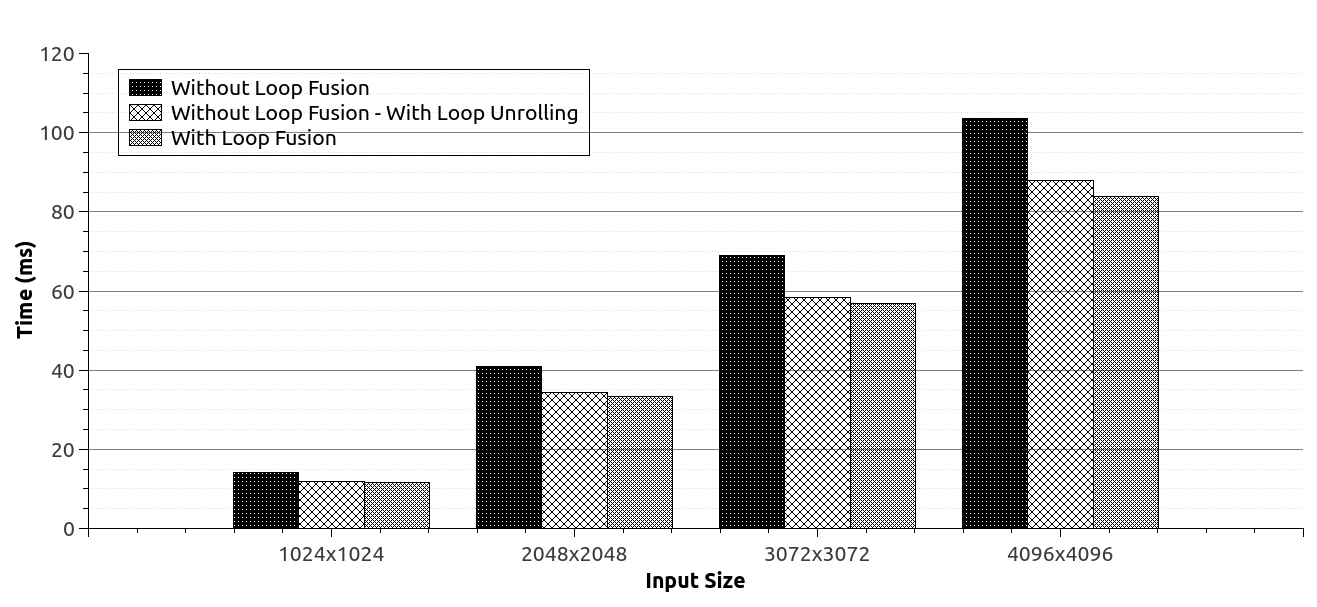}
	\caption{Effect of loop fusion on the CPU with two common data elements.}
	\label{fig:cpu_loop_fusion}
\end{figure}

From the Fig. \ref{fig:cpu_loop_fusion}, a small performance gain can be seen due to loop fusion technique on the CPU. Therefore, it is clear that if the program has more common variables between loops then it will show some performance gain as the effect of loop fusion. For these results, the temporal locality affected for gaining the above performance gain.

\begin{figure}[!t]
	\centering
	\includegraphics[width=3.3in]{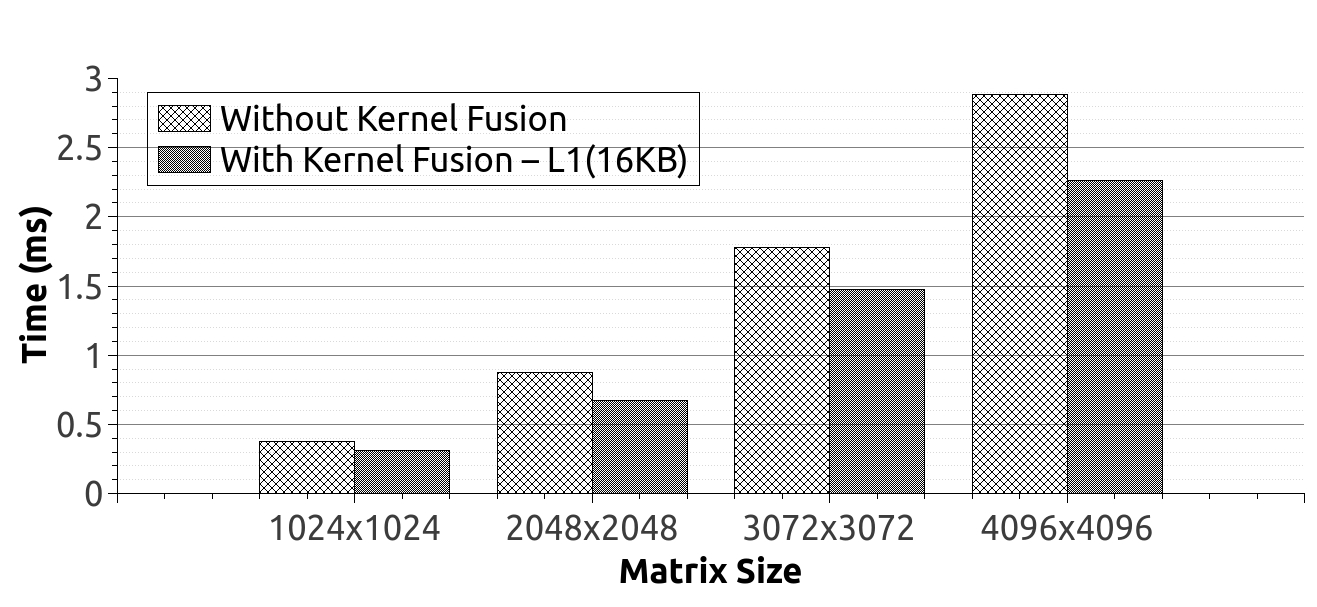}
	\caption{Effect of the kernel fusion on the Fermi GPGPU.}
	\label{fig:gpgpu_kernel_fusion}
\end{figure}

Fig. \ref{fig:gpgpu_kernel_fusion}  shows the main time difference between tests with and without kernel fusion on the GPGPU. Here, it can be identified main two reasons for the performance difference. One reason for taking less time in fused kernel test is executing less number of instruction than separated kernels. Another reason is effective cache usage of combined kernels. Combined kernels increase temporal locality like CPU loop fusion and the result is better performance. Here also, when input size is large, performance gain is also high.

\begin{figure}[!t]
	\centering
	\includegraphics[width=3.3in]{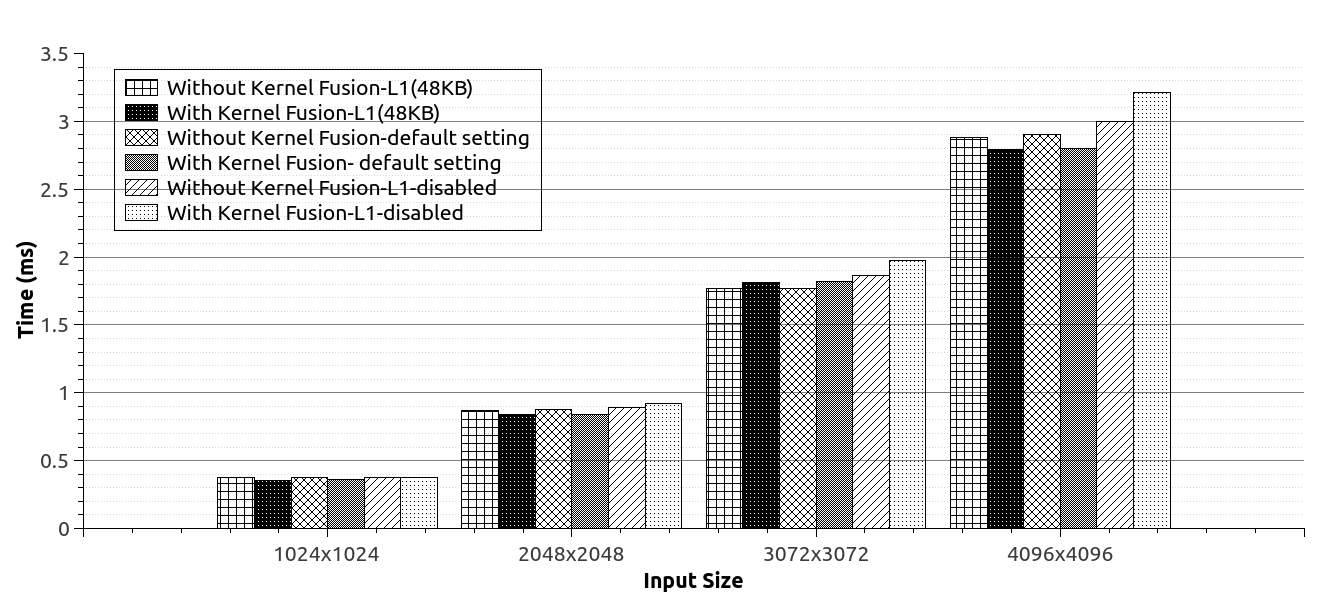}
	\caption{Effect of kernel fusion on the Fermi GPGPU according to various cache configurations.}
	\label{fig:gpgpu_kernel_fusion_2}
\end{figure}

According to the above experimental result set, the main reason for the performance gain of kernel fusion on GPGPU is identified as cache friendly data arrangement of fused kernels of the GPGPU. Fig. \ref{fig:gpgpu_kernel_fusion_2} shows that main factor for the performance is size of the L1 cache because disabled L1 cache shows poor performance than enabled L1 cache.

\subsection{Effect of array merging}
\begin{figure}[!t]
	\centering
	\includegraphics[width=3.3in]{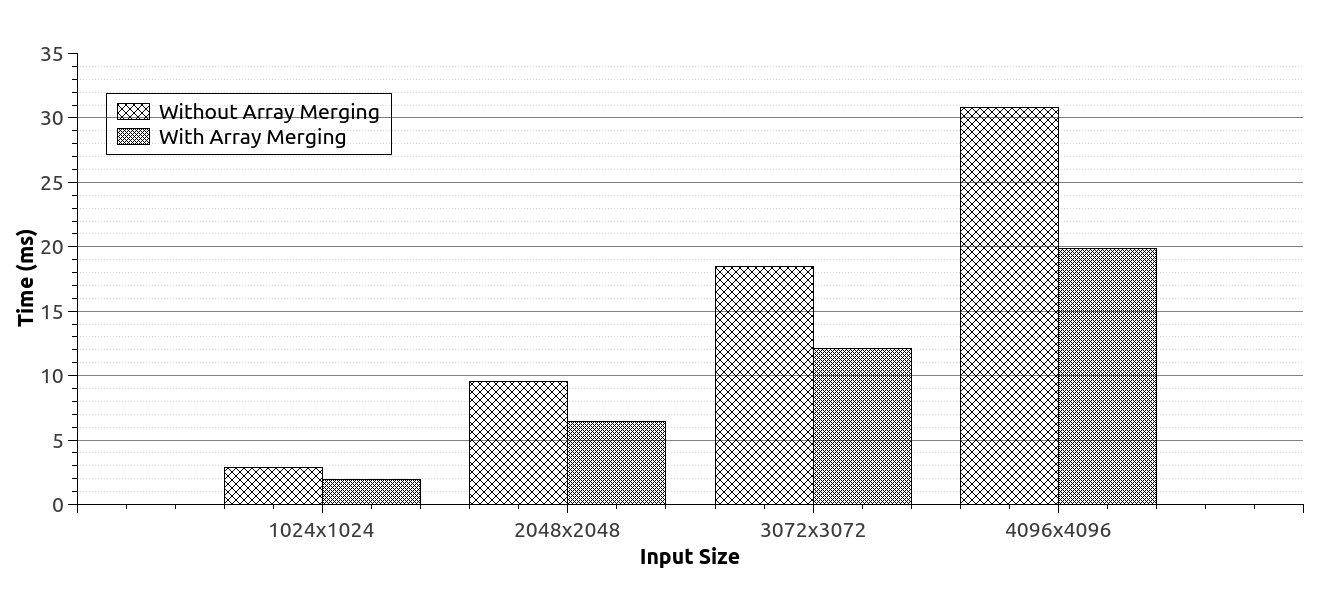}
	\caption{Effect of array merging on the CPU.}
	\label{fig:cpu_array_merging}
\end{figure}

A clear difference can be seen in Fig. \ref{fig:cpu_array_merging}, between the time taken for non-merged and merged arrays in the CPU. Here, merging has increased the effect of spatial locality. As a result, better performance could be gained from this array merging on the CPU.

\begin{figure}[!t]
	\centering
	\includegraphics[width=3.3in]{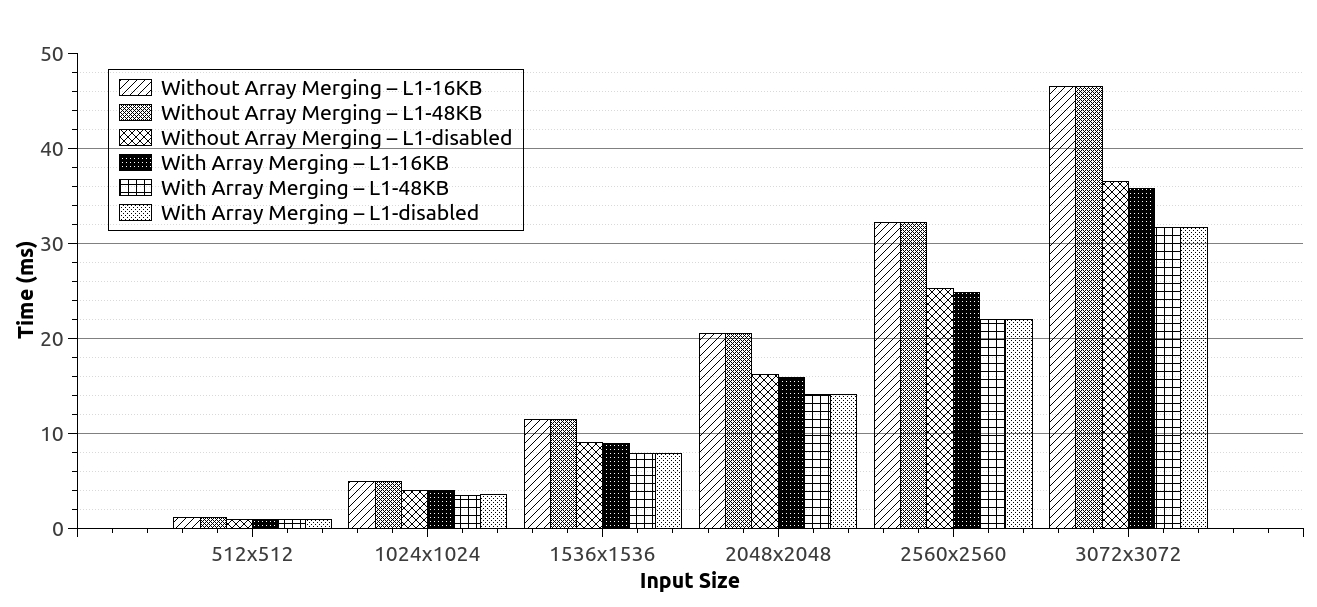}
	\caption{Effect of the array merging on the Fermi GPGPU according to the various cache configurations.}
	\label{fig:gpgpu_array_merging}
\end{figure}

Array merging can be tested on the GPGPU with various configurations like 16KB L1 cache, 48KB L1 cache, and disabled L1 cache. According to Fig. \ref{fig:gpgpu_array_merging}, the amount of L1 cache size does not show any effects to the executions those does not use array merging. Tests without array merging with disabled L1 cache shows some improvement over the enabled L1 cache. The reason for this improvement is, when the L1 cache is disabled, cache line size of L2 cache is converted from 128 bytes to 32 bytes. The result is more cache lines in L2 cache. It reduces evicting more cache lines.

\begin{figure}[!t]
	\centering
	\includegraphics[width=3.3in]{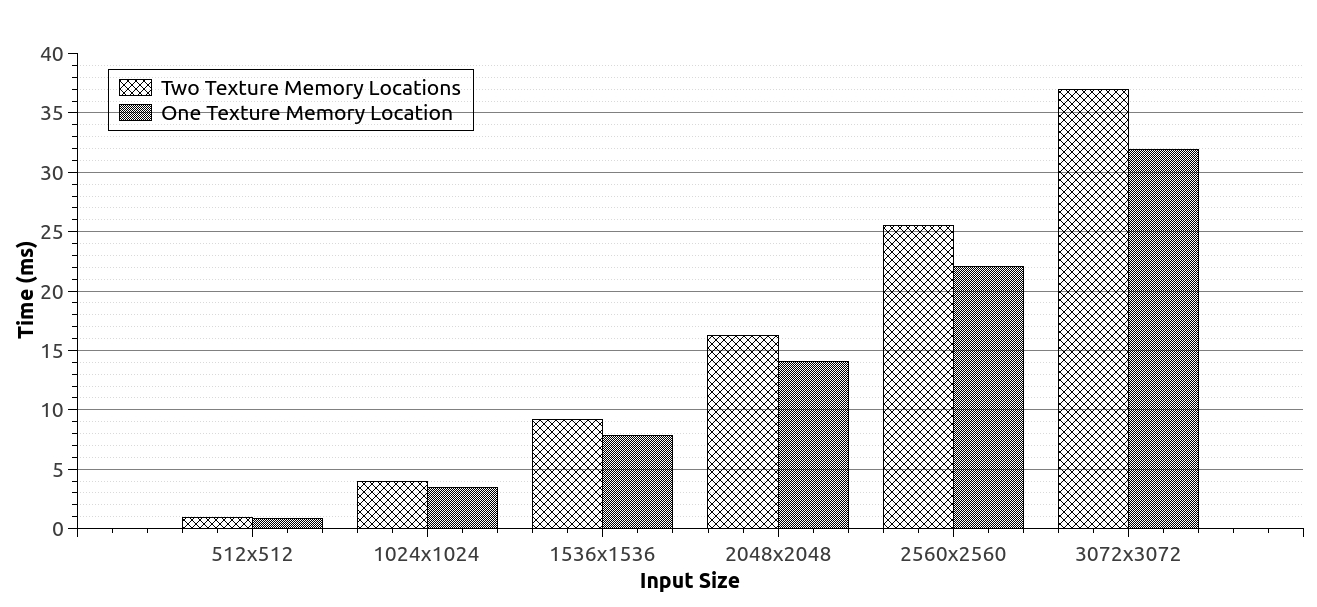}
	\caption{Effect of array merging with texture memory on the Fermi GPGPU.}
	\label{fig:gpgpu_array_merging_2}
\end{figure}

Array merging increases spatial locality. Therefore, merged arrays show better performance than un-merged arrays on GPGPU. When the L1 cache size is 48KB, it shows better performance than 16KB L1 cache. Here also, the reason is less number of evicting cache lines for overwriting new data with the 48KB L1 cache. However, disabled L1 cache also shows equal performance gain like 48KB L1 cache. A number of cache lines are the reason for this performance gain also. It is easy to identify that merged arrays show better performance on texture memory also from Fig. \ref{fig:gpgpu_array_merging_2}. When the input size is being larger and larger, the performance gain of merged arrays also shows better and better performance.

\subsection{Effect of array transpose}

\begin{figure}[!t]
	\centering
	\includegraphics[width=3.3in]{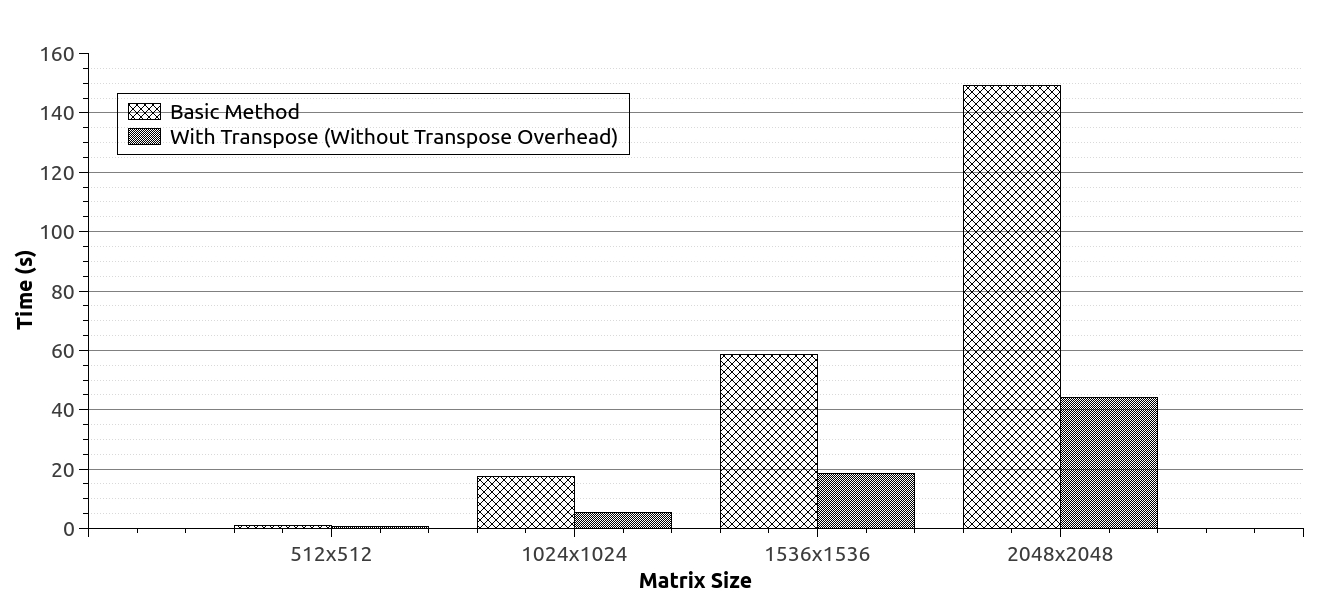}
	\caption{Effect of the array transpose for the matrix multiplication on the CPU.}
	\label{fig:cpu_array_transpose}
\end{figure}

\begin{figure}[!t]
	\centering
	\includegraphics[width=3.3in]{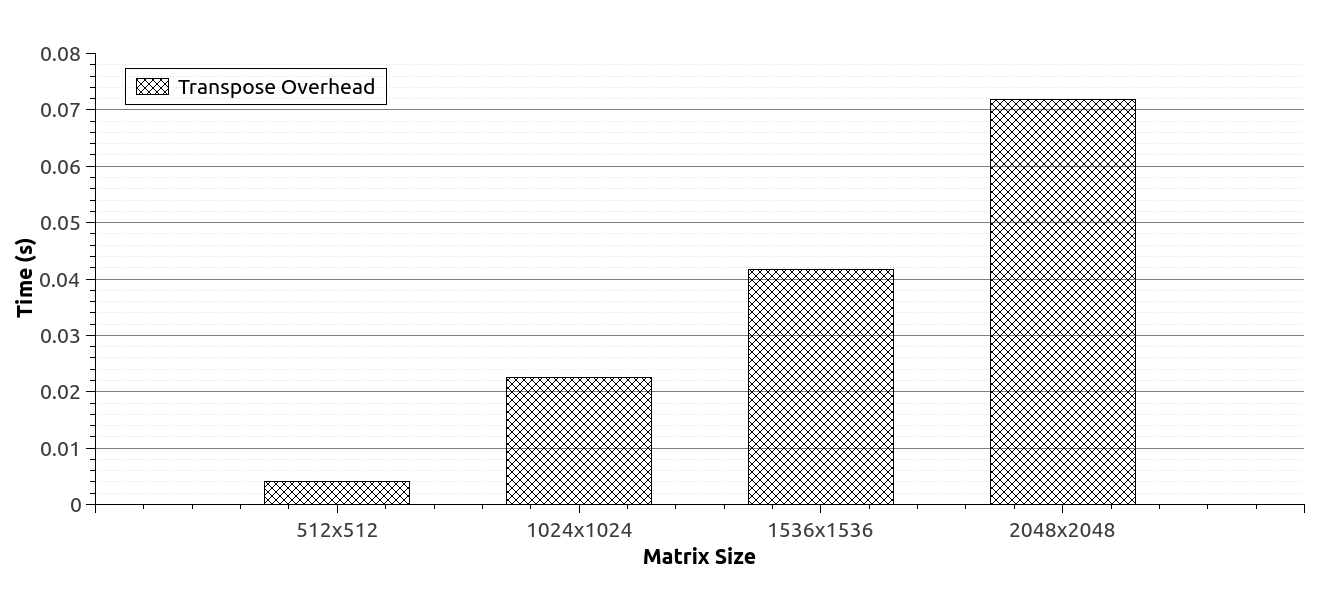}
	\caption{The overhead of array transpose on the CPU.}
	\label{fig:cpu_array_transpose_2}
\end{figure}

The performance difference can be seen from Fig. \ref{fig:cpu_array_transpose}  between normal matrix multiplication and matrix multiplication with array transpose on the CPU. Transposed matrix multiplication shows better  performance gain than normal matrix multiplication in CPU side. The reason is improved spatial locality. In normal matrix multiplication, the second matrix is accessed in N-stride manner. Here N is matrix dimension. It means execution process accesses second matrix elements column by column. Therefore, first element access from one cache line and the second element of the second matrix access from another cache line. If the second matrix is taken as transposed matrix, then it is accessed row by row. The first row is accessed in first iterations and then the second row likewise. Therefore, it increases the spatial locality and result is better performance from the transposed matrix multiplication. However, transposing an array is additional overhead. Therefore, this overhead also was measured during the execution. Time taken for this matrix transpose is graphed in Fig. \ref{fig:cpu_array_transpose_2}. It is clear that overhead of matrix transpose can be ignored. Therefore, performance can be improved in CPU side from a transpose technique and that is suitable for the application.

\begin{figure}[!t]
	\centering
	\includegraphics[width=3.3in]{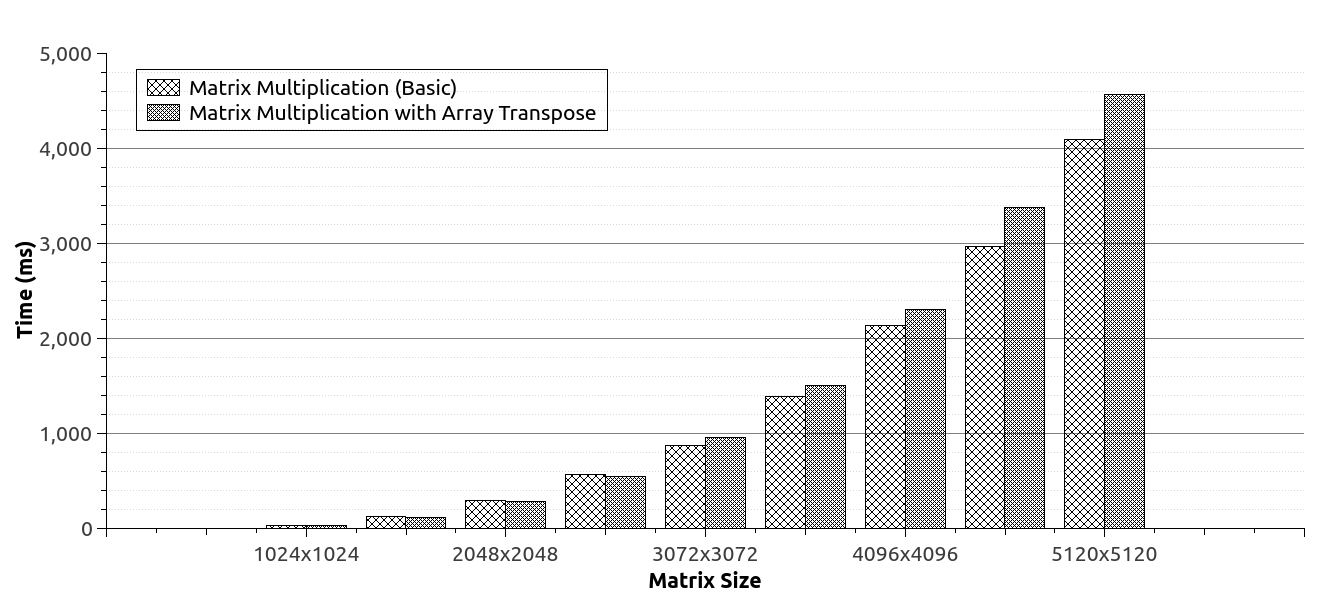}
	\caption{Effect of the array transpose for the matrix multiplication on the Fermi GPGPU.}
	\label{fig:gpgpu_array_transpose}
\end{figure}

Results taken from array transpose tests on the GPGPU are graphed in Fig. \ref{fig:gpgpu_array_transpose}. According to the previous tests on CPU, better performance is expected after applying array transpose on GPGPU also. However, it shows different results from expected outcome. It gives the little gain of performance for less number of matrix sizes with array transpose. Then it shows poor performance for array transpose with large data sets. The reason behind this is, parallel accessing of data elements of a matrix increases usage of number of cache lines in the GPGPU with array transpose technique.

\section{Conclusions and future work}
\balance
Manually loading data into the shared memory of the GPGPU is the best option for gaining best performance with the blocking technique. When there are restrictions for using shared memory of the GPGPU then allocating maximum L1 cache is the best option for achieving better performance. In GPGPUs, if two kernels have common data elements, then it shows better performance for kernel fusion. The reason is the temporal locality of the memory access. Fused and non-fused kernels have a different number of threads. However, the effect of a number of threads to the performance is negligible because GPGPU tests without common data elements between two kernels didn't show performance difference like GPGPU tests with common data elements.

The array merging improves the performance of overall memory access on CPUs as well as GPGPUs. When the L1 cache size is large on GPGPUs, it shows better performance because the cache has enough space to handle memory accesses without evicting more cache lines. However, if the cache size is not enough, then more cache replacements occur. In a such a case, it is better to disable L1 cache and access memory through the L2 cache only. If we are using two texture memory locations, then it is better to merge those arrays to compatible with 2D spatial locality. Transposing arrays, specially 2D arrays are not a good option in GPGPU. The transposed 2D arrays within the GPGPU increases the number of memory accesses through the cache memory structure because it accesses the device memory in parallel using many threads. Therefore, transposed 2D array should be handled within the GPGPU very carefully because it completely depends on the memory access pattern of the program. However, we cannot guarantee the validity of these findings on other GPGPU architectures rather than Fermi architecture.
\balance

\bibliographystyle{IEEEtran}

\bibliography{IEEEabrv,paper1}

\end{document}